\documentclass[
    ,final            % use final for the camera ready runs
  ]
  {aipproc}

\usepackage{psfrag,amsmath,epsfig,a4,latexsym,axodraw}

\layoutstyle{8x11double}

\newcommand{\Eqns}[2]{Eqs.~(\ref{#1}) and (\ref{#2})}

\def\ghat{{\hat{g}}}
\def\gtilde{{\tilde{g}}}
\def\gbar{{\bar{g}}}

\def\ath{{\breve{a}}}

\def\DRED{{\scshape dred}}

\def\HV{{\scshape hv}}
\def\FDH{{\scshape fdh}}
\def\DR{{\scshape dr}}
\def\CDR{{\scshape cdr}}

\def\RS{{\scshape rs}}

\def\mDRED{{\rm\scriptscriptstyle DRED}}
\def\mCDR{{\rm\scriptscriptstyle CDR}}
\def\mHV{{\rm \scriptscriptstyle HV}}
\def\mFDH{{\rm \scriptscriptstyle FDH}}
\def\mRS{{\rm \scriptscriptstyle RS}}

\def\eps{\epsilon}

\begin{document}

\title{Dimensional Reduction and Hadronic Processes}

\classification{ 11.10.Gh, 11.15.-q, 12.38.Bx, 12.60.Jv}

\keywords{}
\author{Adrian Signer}{
  address={Institute for Particle Physics Phenomenology,
Durham University,
Durham DH1~3LE, UK}
}

\author{Dominik St\"ockinger}{
  address={Institut f\"ur Kern- und Teilchenphysik,
TU Dresden, D-01062 Dresden, Germany}
}

\begin{abstract}
We consider the application of regularization by dimensional reduction to NLO
corrections of hadronic processes. The general collinear
singularity structure is discussed, the origin of the
regularization-scheme dependence is identified and transition rules to
other regularization schemes are derived.
\end{abstract}

\maketitle

%%%%%%%%%%%%%%%%%%%%%%%%%%%%%%%%%%%%%%%%%%%%
%% MAINMATTER
%%%%%%%%%%%%%%%%%%%%%%%%%%%%%%%%%%%%%%%%%%%%

\section{Introduction}

The LHC era necessitates the computation of next-to-leading order (NLO)
predictions in the standard model and extensions such as
supersymmetry. One element of such 
computations is the choice of a regularization scheme. In particular,
in supersymmetric theories, regularization by dimensional
reduction \cite{Siegel79} is often advantageous compared to
dimensional regularization, which breaks supersymmetry.
In recent years, progress on the understanding of dimensional
reduction has been achieved in three directions. A consistent
definition and proofs of various supersymmetry relations have been
obtained \cite{DS05,HollikDS05}; multiloop applications have been
pioneered \cite{SteinhauserKantPapers}, and the
factorization problem found in Refs.~\cite{Beenakker:1988bq,
Beenakker:1996dw, Smith:2004ck} has been
resolved \cite{Signer:2005iu}.

In these proceedings we present our study \cite{SignerDSnew}, where we
reconsider the factorization problem in a more general context. We
consider real and virtual NLO QCD corrections to
arbitrary hadronic $2\to (n-2)$ processes with massless or massive
partons. We discuss the infrared singularity structure and the
associated regularization-scheme dependence of all these corrections,
provide transition rules between the schemes and show that all
singularities factorize. In this way we show that the framework of
dimensional reduction is completely consistent with factorization, and
we show how this scheme can be used to compute hadronic processes in
practice.

\section{Four regularization schemes}

In a first step we need to precisely define the
regularization schemes.
As it turns out, in the literature two different
versions of dimensional reduction with different factorization
behaviour have been used, and it is crucial to distinguish between
them. 

In both dimensional regularization and dimensional reduction,
space-time and momenta are continued from $4$ to $D=4-2\epsilon$
dimensions. Gluon fields (and other vector fields) are basically
treated as $D$-dimensional in dimensional regularization and
$4$-dimensional in dimensional reduction. In the consistent definition
of dimensional reduction \cite{DS05} three spaces are distinguished:
the original $4$-dimensional Minkowski space, the $D$-dimensional
space for regularized momenta and space-time coordinates, and a
formally $4$-dimensional space for the regularized gluon fields. The
associated metric tensors are denoted as $\gbar^{\mu\nu}$,
$\ghat^{\mu\nu}$, and $g^{\mu\nu}$, respectively.  The dimensionalities of the spaces are
expressed by the following equations:
\begin{align}
 g^{\mu\nu}g_{\mu\nu} &= 4,&
 \ghat^{\mu\nu}\ghat_{\mu\nu} &= D,&
 \gbar^{\mu\nu}\gbar_{\mu\nu} &= 4.
 \label{ggDef}
\end{align} 
The following projection relations express the hierarchical structure
of the three spaces:
\begin{align} 
g^{\mu\nu}\ghat_\nu{}^\rho&=\ghat^{\mu\rho},&
g^{\mu\nu}\gbar_\nu{}^\rho&=\gbar^{\mu\rho},&
\ghat^{\mu\nu}\gbar_\nu{}^\rho&=\gbar^{\mu\rho}.
\label{ggProjections}
\end{align}
It is useful to introduce the orthogonal complement to the
$D$-dimensional space. This is
a $4-D=2\epsilon$-dimensional space with metric tensor
$\gtilde^{\mu\nu}$, which satisfies
\begin{align}
\label{gghatgtilde}
g^{\mu\nu}&=\ghat^{\mu\nu}+\gtilde^{\mu\nu},&
\gtilde^{\mu\nu}\gtilde_{\mu\nu}&=2\epsilon,\\
g^{\mu\nu}\gtilde_{\nu}{}^\rho&=\gtilde^{\mu\rho},&
\ghat^{\mu\nu}\gtilde_{\nu}{}^\rho&=0,&
\gbar^{\mu\nu}\gtilde_{\nu}{}^\rho&=0.
\end{align}

It is not strictly necessary to regularize all gluons. Gluons which
appear inside a closed loop or inside a singular region (soft or
collinear) of a phase space integral are called ``internal'', and
these need to be regularized. All other gluons are called
``external''; they require no regularization.
As a result it is
possible to distinguish two versions \CDR\ and \HV\ of dimensional regularization and
two versions of dimensional reduction \DRED\ and \FDH, depending on whether external
gluons are treated in the same way as internal ones or not.
The following table defines the four schemes by specifying which
metric tensor is to be used for the gluons in the gluon
propagators/polarization sums:
\begin{center}
\begin{tabular}{l|cccc}
\hline
&\CDR&\HV&\FDH&\DRED\\
\hline
internal gluon&$\ghat^{\mu\nu}$&$\ghat^{\mu\nu}$&
$g^{\mu\nu}$&$g^{\mu\nu}$\\
 external gluon&$\ghat^{\mu\nu}$&$\gbar^{\mu\nu}$&
$\gbar^{\mu\nu}$&$g^{\mu\nu}$
\\
\hline
\end{tabular}
\end{center}

Note that the \FDH\ version of dimensional reduction \cite{FDHrefs}
has been denoted as \DR\ e.g.\
in Refs.~\cite{Kunszt:1993sd,Catani:1996pk,Catani:2000ef} (for the
one-loop equivalence see  e.g.\ 
Refs.~\cite{Kunszt:1993sd,Bern:1996je}). The scheme \DRED\ is the one
defined in e.g.~\cite{Siegel79,CJN80,DS05}.

\section{Infrared structure and factorization in the four schemes}

\subsection{Factorization problem}

In Refs.\ \cite{Beenakker:1988bq,Smith:2004ck} an apparent
non-factorizing behaviour
of \DRED\ has been found in the real corrections to the process
 $gg\to t\bar{t}g$. The reason for this apparent problem has
been identified in Ref.\ \cite{Signer:2005iu}. On the regularized
level in \DRED, the splitting gluon cannot be treated as a single,
formally 4-dimensional gluon, but it should be decomposed into its
$D$- and $(4-D)$-dimensional parts, according to
$g^{\mu\nu}=\ghat^{\mu\nu}+\gtilde^{\mu\nu}$. The $\ghat$-part behaves
as a $D$-dimensional gauge field, while $\gtilde$ behaves as $2\eps$
scalar fields, the $\eps$-scalars. If this decomposition is taken into
account factorization holds in \DRED\ just as expected in a theory
with two different partons $\ghat$ and $\gtilde$.

\subsection{Splittings}

The regularization-scheme (\RS) dependence of real and virtual
corrections to arbitrary processes is related to ultraviolet, soft,
and collinear singularities. In these proceedings we only take into
account collinear singularities, assuming fully renormalized and thus
ultraviolet finite amplitudes and noting that the soft singularities
only lead to a trivial \RS\ dependence. The collinear singularities of
real corrections factorize into products of lowest-order matrix
elements and splitting functions, and the associated \RS\ dependence
is expressed in terms of the \RS\ dependence of the splitting
functions.

The splitting functions $P^{<\, \mRS}_{i^*\to jk}(z)$ describe the
splitting of a parton $i$, which is slightly off-shell, into collinear
partons $j$, $k$, where the momenta of $j$, $k$ are given by $z$ and
$(1-z)$ times the momentum of $i$ (for $z<1$). Most interesting for our purposes
are the splitting functions involving gluons, which are \RS\ dependent
owing to the different gluon prescriptions.

\begin{figure}[tb]
\scalebox{.9}{
\begin{picture}(400,100)
\SetOffset(20,20)
\Text(30,-20)[l]{\CDR}
\Text(25,28)[lb]{$\ghat$}
\Text(65,55)[lb]{$\ghat$}
\Text(74,41)[lb]{$\ghat$}
\Gluon(15,15)(44.6,30){2}{4}
\Gluon(44.6,30)(65.9,51.2){2}{4}
\Gluon(44.6,30)(73.6,37.8){2}{4}
\Line(10,10)(37.2,-2.7)
\Line(10,10)(39.9,7.4)
\Line(10,10)(29.3,-13.)
\GOval(10,10)(10,10)(0){0.8}
\SetOffset(120,20)
\Text(30,-20)[l]{\HV}
\Text(25,28)[lb]{$\bar{g}$}
\Text(65,55)[lb]{$\ghat$}
\Text(74,41)[lb]{$\ghat$}
\Gluon(15,15)(44.6,30){2}{4}
\Gluon(44.6,30)(65.9,51.2){2}{4}
\Gluon(44.6,30)(73.6,37.8){2}{4}
\Line(10,10)(37.2,-2.7)
\Line(10,10)(39.9,7.4)
\Line(10,10)(29.3,-13.)
\GOval(10,10)(10,10)(0){0.8}
\SetOffset(220,20)
\Text(30,-20)[l]{\FDH}
\Text(25,28)[lb]{$\bar g$}
\Text(65,55)[lb]{$g$}
\Text(74,41)[lb]{$g$}
\Gluon(15,15)(44.6,30){2}{4}
\Gluon(44.6,30)(65.9,51.2){2}{4}
\Gluon(44.6,30)(73.6,37.8){2}{4}
\Line(10,10)(37.2,-2.7)
\Line(10,10)(39.9,7.4)
\Line(10,10)(29.3,-13.)
\GOval(10,10)(10,10)(0){0.8}
\SetOffset(320,20)
\Text(30,-20)[l]{\DRED}
\Text(25,28)[lb]{$g$}
\Text(65,55)[lb]{$g$}
\Text(74,41)[lb]{$g$}
\Gluon(15,15)(44.6,30){2}{4}
\Gluon(44.6,30)(65.9,51.2){2}{4}
\Gluon(44.6,30)(73.6,37.8){2}{4}
\Line(10,10)(37.2,-2.7)
\Line(10,10)(39.9,7.4)
\Line(10,10)(29.3,-13.)
\GOval(10,10)(10,10)(0){0.8}
\end{picture}
}
\caption{ Gluon splitting into two collinear
  gluons in the four schemes, indicating the appropriate treatment of
  each gluon.\label{fig:splitggg} }
\end{figure}
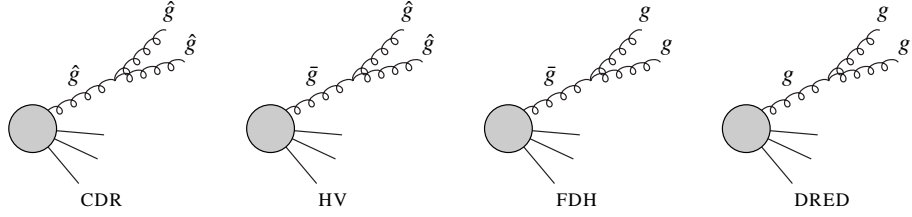
Figure~\ref{fig:splitggg} shows the four different gluon prescriptions
for $P^{<\, \mRS}_{g^*\to gg}$. According to the definition given
above, the two collinear gluons $j$ and $k$ are treated as
``internal'', and the virtual gluon $i$ as ``external''. 

In order to understand the \RS\ dependence it should first be noted
that the projection from $\ghat$ onto $\gbar$ does not change the
result of the splitting functions since $\gbar$ is simply a part of
the $D$-dimensional gauge field and thus behaves in the same way as
$\ghat$. Then Fig.\ \ref{fig:splitggg} shows how the results change in
going from \CDR\ to \HV, \FDH, and \DRED:
\begin{align}
\mbox{\CDR:}&&P^{<\, \mCDR}_{g^*\to gg} &=
P^{<\, \mDRED}_{\ghat^*\to\ghat\ghat}\\
\mbox{\HV:}&&P^{<\, \mHV}_{g^*\to gg} &=
P^{<\, \mDRED}_{\ghat^*\to\ghat\ghat}\\
\mbox{\FDH:}&&P^{<\, \mFDH}_{g^*\to gg} &=
P^{<\, \mDRED}_{\ghat^*\to\ghat\ghat}+P^{<\, \mDRED}_{\ghat^*\to\gtilde\gtilde}\\
\mbox{\DRED:}&&P^{<\, \mDRED}_{\ghat^*\to gg} &=
P^{<\, \mDRED}_{\ghat^*\to\ghat\ghat}+P^{<\, \mDRED}_{\ghat^*\to\gtilde\gtilde}\\
&&P^{<\, \mDRED}_{\gtilde^*\to gg} &=
P^{<\, \mDRED}_{\gtilde^*\to\gtilde\ghat}+P^{<\, \mDRED}_{\gtilde^*\to\ghat\gtilde}
\end{align}
In words, in \FDH\ there is a new final state, $\gtilde\gtilde$, which
modifies the splitting function, and in \DRED\ there is a new initial
state of the splitting, $\gtilde$, which gives rise to an independent
splitting function. This reflects the discussion of the previous
subsection. Splitting functions involving quarks are related
in a similar way.

The splitting functions $P^{<\,\mRS}_{i^*\to\rm anything}$ defined for
$z<1$ give rise to \RS\ dependent constants $\gamma_\mRS(i)$
\begin{align}
\gamma_\mRS(i)&=-\sum_{k,l}\int_0^1 dz \ z
\frac{(1-z)}{(1-z)_+} P_{i^*\to k l}^{<\, \mRS}(z)
.
% \\
%&=-\sum_{k,l}\int_0^1 dz \, \Theta\left(z-\frac{1}{2}\right) 
%\frac{(1-z)}{(1-z)_+} P_{i^*\to kl}^{<\, \mRS}(z)\, .
%\nonumber
\end{align}
Via unitarity, these constants $\gamma_{\mRS}(i)$ are the origin of the \RS\
dependence of the virtual
corrections \cite{Kunszt:1993sd,Catani:1996pk}. 
As above, the \RS\ dependence of these constants is easily understood:
$\gamma_\mCDR(i)=\gamma_\mHV(i)$, the differences
$\gamma_\mFDH(i)-\gamma_\mHV(i)$ are non-zero due to the possible
splittings $\ghat\to\gtilde\gtilde$ and $q\to q\gtilde$, and in \DRED\
there is a new constant $\gamma_\mDRED(\gtilde)$, but otherwise the
results in \DRED\ and \FDH\ are the same,
$\gamma_\mDRED(\ghat)=\gamma_\mFDH(g)$ and $\gamma_\mDRED(q)=\gamma_\mFDH(q)$.
The results in \DRED\ are new, while the results in the other schemes
have already been obtained in
Refs.~\cite{Kunszt:1993sd,Catani:1996pk}.

\subsection{Results for squared matrix elements}

The collinear singularities of real corrections in $\mRS*\in\{\mCDR, \mHV,
 \mFDH\}$ are well known. In these schemes, if two outgoing partons
$\bar{a}_k$ and $\bar{a}_l$ 
become collinear, the squared matrix element for a process involving
$\bar{a}_k$ and $\bar{a}_l$ satisfies 
\begin{eqnarray}
\lefteqn{{\cal M}^{(0)}_{\mRS*}
    (a_1,a_2;\ldots \bar a_l(p_l)\ldots \bar a_k(p_k)\ldots ) 
\stackrel{p_k\| p_l}{=} \frac{2\, g_s^2}{s_{kl}}\times} 
\label{CollLim} \\
&& 
P_{{(kl)}^*\to k l}^{<\, \mRS*}(z)\, 
{\cal M}^{(0)}_{\mRS*}(a_1,a_2;\ldots a_{(kl)}(p_k+p_l)\ldots ) .
\nonumber
\end{eqnarray}
Here $(kl)$ denotes the (uniquely determined) flavour of the splitting
$(kl)\to kl$. Our new result for \DRED\ can be written in the same
form,
\begin{eqnarray}
\lefteqn{{\cal M}^{(0)}_{\mDRED}
    (a_1,a_2;\ldots \bar a_l(p_l)\ldots \bar
  a_k(p_k)\ldots ) 
\stackrel{p_k\| p_l}{=}   \frac{2\, g_s^2}{s_{kl}}\times}
\label{CollLimDRED} \\
&& \sum_{\ath_{(kl)}}
P_{{(kl)}^*\to k l}^{<\, \mDRED}(z)\, 
{\cal M}^{(0)}_{\mDRED}(a_1,a_2;\ldots
\ath_{(kl)}(p_k+p_l)\ldots ) .
\nonumber
\end{eqnarray}
Here the split $g=\ghat+\gtilde$ becomes
essential and therefore there is a sum over all possible splittings
$\sum_{\ath_{(kl)}}$, where $\ath_{(kl)}\in\{\ghat,\gtilde\}$ if
$(kl)$ is a gluon, and $\ath_{(kl)}\in\{q\}$ if $(kl)$ is a quark. 

The collinear singularities of virtual corrections in $\mRS*\ne\mDRED$
are given by \cite{Kunszt:1994mc,Catani:2000ef}
\begin{eqnarray}
\label{Mvirtual*}
&&{\cal M}^{(1)}_{\mRS*}(a_1\ldots a_n) =
\frac{\alpha_s}{2\pi} \times\\
&&\quad   
\sum_{i} {\cal M}^{(0)}_{\mRS*}(a_1\ldots a_n) \left(
    - \frac{1}{\epsilon}  \gamma_{\mRS*}(a_i) \right)
+\ldots
 \nonumber
%\sum_{i, j} {\cal V}(i,j)\, 
%     {\cal M}^{ij}_{\mRS*}(a_1\ldots a_n)
%   + {\cal M}_\mNS^{(1)}(a_1\ldots a_n) \Bigg]  ,
\end{eqnarray}
where ${\cal M}^{(1)}$ denotes the fully renormalized one-loop squared
matrix element, the dots denote finite terms and soft singularities,
and the sum over $i$ is over all external legs.
In \DRED, the result has a similar form,
\begin{eqnarray}
\label{MvirtualFull}
&&{\cal M}^{(1)}_{\mDRED}(a_1\ldots a_n)
 = 
\frac{\alpha_s}{2\pi}  \times\\
&&\quad
\sum_i\sum_{\ath_i}\, 
 {\cal M}^{(0)}_{\mDRED}(a_1\ldots \ath_i\ldots a_n) \left(
    - \frac{1}{\epsilon} \gamma_{\mDRED}(\ath_i) \right)
+\ldots
\nonumber
\end{eqnarray}
Again, the only difference is the additional sum over the two
possibilities $\ath_{(kl)}\in\{\ghat,\gtilde\}$ if $(kl)$ is a gluon.

From \Eqns{Mvirtual*}{MvirtualFull} and the explicit results for the
$\gamma(i)$ \cite{SignerDSnew} one can obtain explicit 
rules for translating the results in one scheme into results in any of
the other schemes.

\section{Conclusions for practical applications}

Based on the main results on the singularity structure in \DRED\
discussed above, one can derive two further crucial
consequences \cite{SignerDSnew}: (1) it is possible to realize the
$\overline{\mbox{MS}}$-factorization scheme, even if e.g.\ \DRED\ is
used, and (2) even in \DRED\ no parton distribution functions for the
unphysical $\eps$-scalars are required. Hence, even in \DRED\ the
standard, $\overline{\mbox{MS}}$-PDF can be used.  

There are explicit, simple rules on how to transform the various parts
(real and virtual corrections, collinear counterterm) from \DRED\ to
other \RS\ or vice versa. It is thus possible to use different \RS\
for different parts of an NLO computation, depending on which is most
practical. Since \DRED\ is better compatible with supersymmetry, it
might be a simplification to apply \DRED\ in particular to the
computation of virtual corrections.



\bibliographystyle{aipproc}   % if natbib is available


%%%%%%%%%%%%%%%%%%%%%%%%%%%%%%%%%%%%%%%%%%%
%% The following lines show an example how to produce a bibliography
%% without the help of the BibTeX program. This could be used instead
%% of the above.
%%%%%%%%%%%%%%%%%%%%%%%%%%%%%%%%%%%%%%%%%%%

\end{document}